\documentclass[12pt,fleqn]{iopart}

\usepackage{iopams} 
\usepackage{harvard} 
\usepackage{graphicx}

\expandafter\let\csname equation*\endcsname\relax
\expandafter\let\csname endequation*\endcsname\relax 
\usepackage{amsmath}

\def\zquiv{z_{\rm quiv}}
\def\ti{t'}\def\Fi{F'}
\def\prhoi{p_{\!\rho}'}

\begin{document}

\title{Energy bunching in soft recollisions revealed with long-wavelength few-cycle pulses}

\author{Alexander K\"astner$^{1}$, Ulf Saalmann$^{1,2}$, and Jan M.\ Rost$^{1,2}$}

\address{$^1$Max Planck Institute for the Physics of Complex Systems\\
   N\"othnitzer Stra{\ss}e 38, 01187 Dresden, Germany\\
   $^2$Max Planck Advanced Study Group at CFEL\\
   Luruper Chaussee 149, 22761 Hamburg, Germany}
\ead{us@pks.mpg.de}
\begin{abstract}\noindent
Soft recollisions are laser-driven distant collisions of an electron with its parent ion.
Such collisions may cause an energy bunching, since electrons with different initial drift momenta
can acquire impacts, which exactly counterbalance these differences.
The bunching generates a series of peaks in the photo-electron spectrum.
We will show that this series could be uncovered peak-by-peak experimentally by means of phase-stabilized few-cycle pulses with increasing duration.
\end{abstract}

\pacs{34.80.Qb,32.80.Rm,32.80.Wr,32.80.Fb}
\maketitle

\section{Introduction}
In addition to  high-energy phenomena such as high-harmonic generation and above-threshold ionization, the interaction of intense long-wavelength (a few $\mu$m) laser pulses with atoms also leads to an interesting low-energy structure (LES) in the photo-electron spectrum close to threshold \cite{blca+09}, which could be reproduced with numerical spectra, quantum mechanically \cite{blca+09} as well as classically \cite{quli+09,liha10,ledi+11}. Recently, soft recollisions, where the electron turns around, i.e., goes through momentum zero, at some distance from the ion, have been uncovered as the origin of the LES with the analytical prediction that a series of low-energy peaks should exist \cite{kasa+11}. In addition, the number of peaks increases with the number of laser cycles contained in the full pulse.
However, those peaks are difficult to access experimentally since they sit on top of a large background signal which strongly varies at low energies towards threshold. Moreover, there may be in addition low-energy features depending specifically on individual potential properties.

These difficulties could be overcome if a few-cycle pulse is used for the measurement. In addition, such a pulse should also resolve the generation of the LES peak-by-peak with successively increasing pulse lengths. However, using ultrashort pulses will shift the peak positions compared to the analytically determined ones. Moreover, the carrier-envelope phase (CEP), i.\,e., the phase difference between the maximum of a cycle and the maximum of the envelope \cite{pagr+01,baud+03}, becomes relevant for the peak positions. 
For these reasons we provide in the following numerical photo-electron spectra for such pulses as a guideline for future experiments.  

\section{Effect of soft recollisions on long-wavelength photo-electron spectra}
\begin{figure}[t]
\begin{center}
\includegraphics[scale=.5]{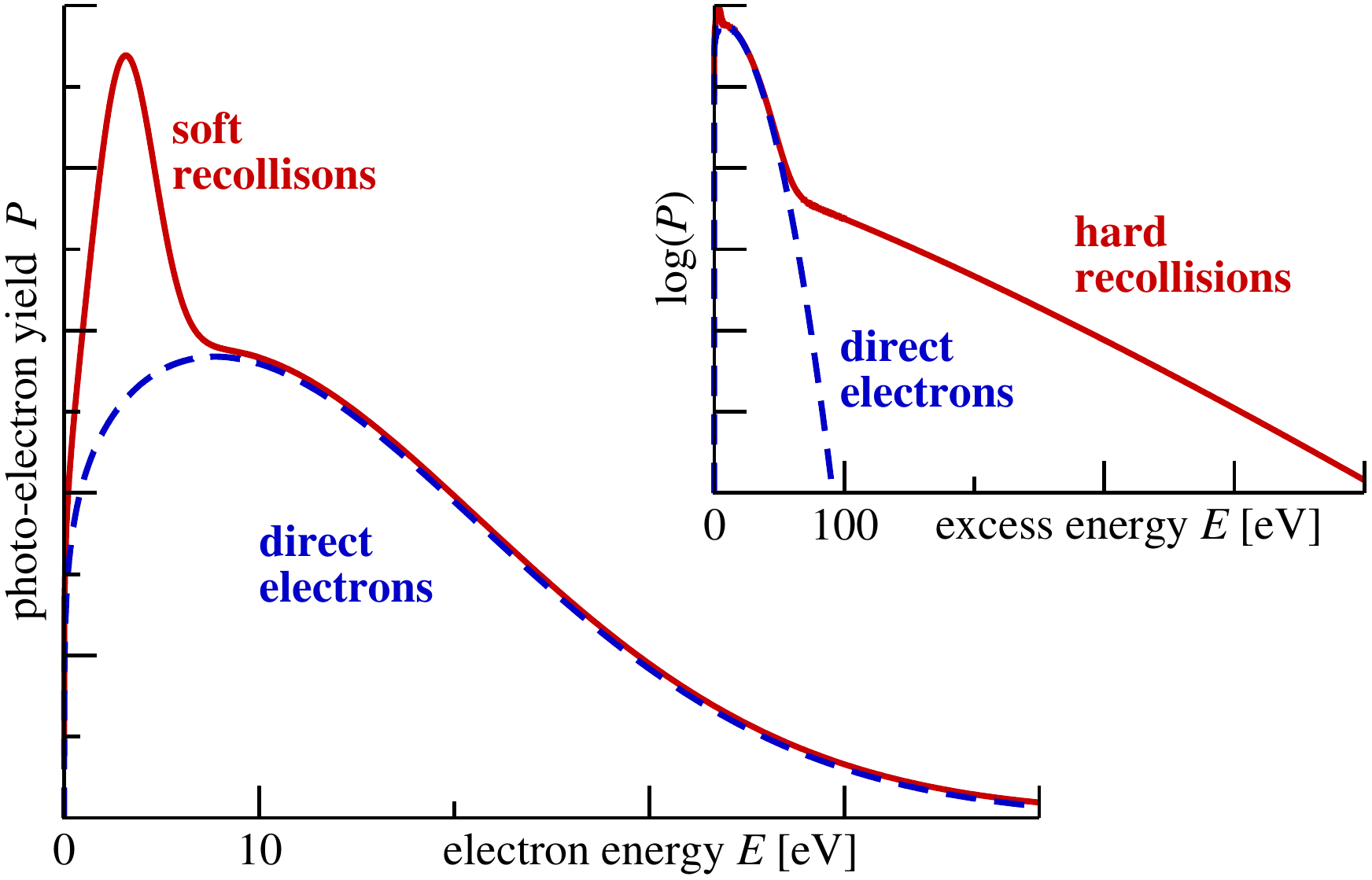}
\caption{Schematic photo-electron spectrum $P(E)$ of atoms in intense long-wavelength laser pulses. The blue dashed line indicates direct photo-ionization, the solid red line shows the spectrum including recollisions.
At low energies soft recollisions induce a pronounced structure, the so-called LES, which is shown on a linear scale (left graph).
Hard recollisions extend the spectrum to higher energies, which becomes visible only on a logarithmic scale (right).}
\label{fig:spectrum_sketch}
\end{center}
\end{figure}%
Before addressing few-cycle pulses directly,
we briefly review the origin of the series of peaks \cite{kasa+11} which we have identified as the LES and discuss how these
peaks add to the other characteristics of the photo-electron spectrum generated with intense pulses.
Typically, upon interaction with a strong oscillating laser field $F(t) = F_{0}\cos(\omega t)$, characterized by peak amplitude $F_{0}$ and frequency $\omega$, the bound electrons leave the ion through tunneling ionization at different phases $\varphi' = \omega t'$ of the laser field\footnote{We denote all initial variables with single prime and all final quantities with a double prime.} with a momentum $p_{\rm direct}(\varphi') = -(F_{0}/\omega)\sin\varphi'+\Delta p$. Hereby $\Delta p$, which accounts for the Coulomb attraction during release, does not depend on $\varphi'$.  All these events are well described by the strong-field approximation (which usually assumes a zero-range potential, with $\Delta p{=}0$), which essentially states that the electrons reaches the detector {\it directly} with the unchanged initial momentum $p_{\rm direct}(\varphi')$. In connection with the $\varphi'$-dependent tunneling probabilities, this leads to the typical well known shape of the spectrum, see blue dashed line in Fig.\,\ref{fig:spectrum_sketch}. For specific initial laser phases $\varphi'$, the electron returns to the nucleus with its center executing a hard recollision by hitting the nucleus head-on. Thereby the electron acquires up to 10$E_{\rm pond}$ with the ponderomotive energy $E_{\rm pond}=F^{2}/(2\omega)^{2}$. This leads to a high-energy component in the photo-electron spectrum albeit of exponentially small magnitude as shown in the right panel of Fig.\,\ref{fig:spectrum_sketch}. 
This phenomenon is known as the plateau in above-threshold ionization \cite{pani+94}.

\begin{figure}[t]
\begin{center}
\includegraphics[scale=0.7]{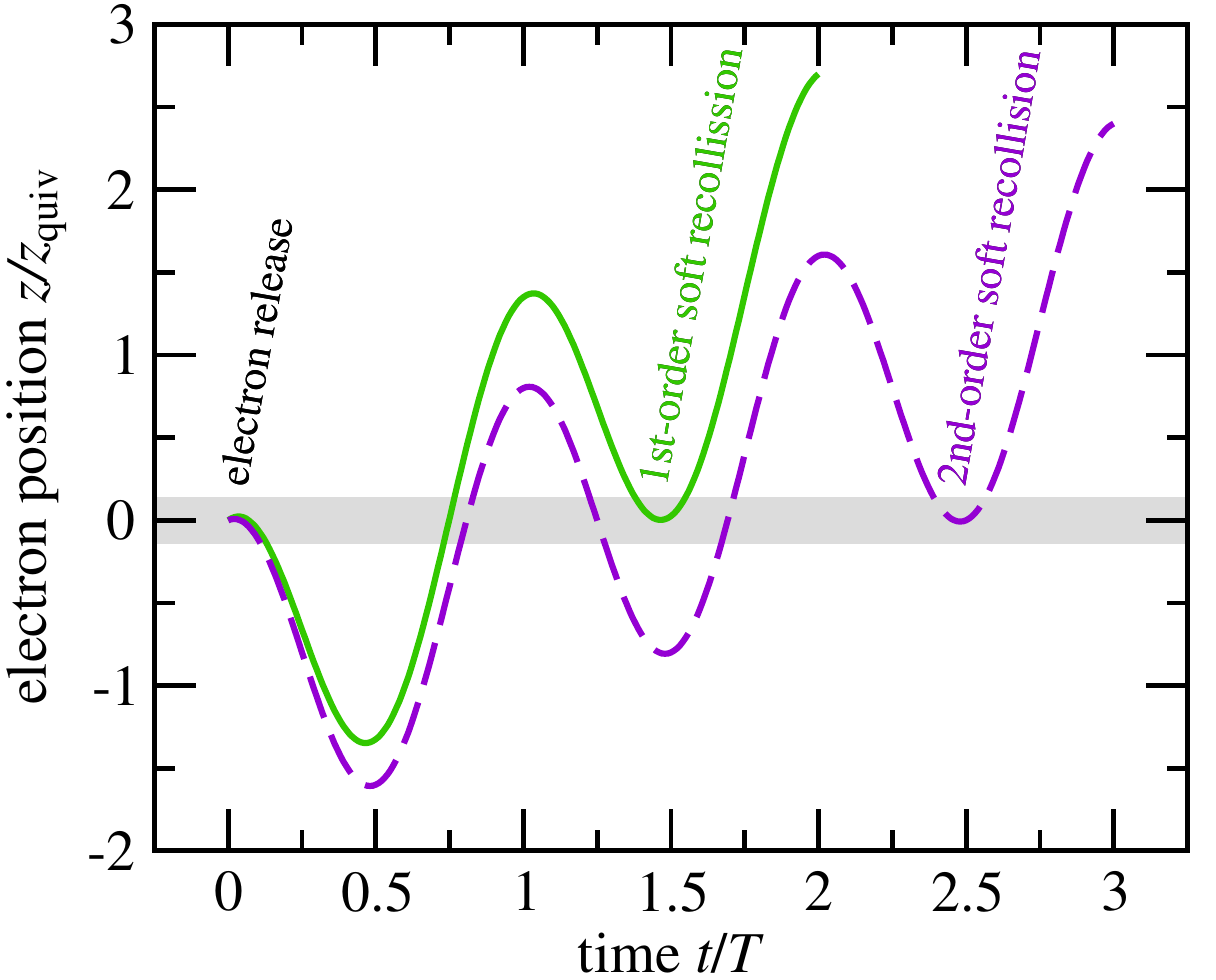}
\caption{Sketch of a 1st-order (solid green line) and 2nd-order (dashed violet line) soft recollision. Shown is electron position $z$ along the laser polarization axis in units of the quiver amplitude $z_{\rm quiv}=F_{0}/\omega^{2}$ as a function of time $t$ in units of the laser period $T$ for an electric field $F(t)=F_{0}\cos(\omega t)$. 
Note that the recollisions occur at finite $\rho$.}
\label{fig:trajectories}
\end{center}
\end{figure}

Complementarily, an electron released around a different specific phase $\varphi'$ of the laser, can be driven aside the nucleus with \emph{minimal\/} kinetic energy in the vicinity of the nucleus. This amounts to a low-energy collision in contrast to the high-energy recollision discussed before and the peaks arising from this soft recollision appear at very low energy in the photo spectrum, see left panel of Fig.\,\ref{fig:spectrum_sketch}. The peaks appear since the soft recollision has a collimation effect on the final electron energy which we have called electron-energy bunching \cite{kasa+11}: The soft recollision can be thought of giving the laser-driven electron a small impact $\delta p$ through the interaction with the ion potential changing the initial momentum $p_\mathrm{final}=p_{\rm direct}(\varphi') +\delta p(\varphi')$.
The impact depends through the turning point on the strong-field trajectory and thereby on the initial phase $\varphi'$.
If the impact $\delta p$ exactly compensates the change of the direct momentum, i.\,e.\ $d(p_{\rm direct}+\delta p)/d\varphi' = 0$, a low-energy peak in the photo-electron spectrum emerges.
Since soft recollisions occur at times $t_{k}=(k{+}1/2)T$, with $T$ the period of the laser, as shown for $k{=}1,2$ in Fig.\,\ref{fig:trajectories}, a series of peaks can emerge, provided the laser pulse is long enough. Obviously, the peak positions are given approximately by \cite{kasa+11}
\begin{equation}
\label{eq:ppeaks}
p_{k} = \frac{F}{\omega}\frac{2}{(2k{+}1)\pi}
\end{equation}
and the corresponding energies read
\begin{equation}
\label{eq:epeaks}
E_{k} = \frac{8}{(2k{+}1)^{2}\pi^{2}}E_{\rm pond}
\end{equation}
The experimentally observed peaks \cite{blca+09} are for $k=1$, i.e.,
$E\approx E_{\rm pond}/10$.
As we will see below, a pulse envelope can shift the peak position with respect to Eq.\,(\ref{eq:epeaks}), where an infinite pulse was assumed.

\section{Computation of the photo-electron dynamics}
\subsection{Hamiltonian}
To obtain a photo-electron spectrum from illumination of atoms with few-cycle pulses we propagate classically electrons in the field of a laser pulse, linearly polarized along $\hat z$, and in the ionic Coulomb potential, after the electrons
have been released according to a tunnel probability, described below. The Hamiltonian of the electron in cylindrical coordinates $\rho$ and $z$ reads
\begin{equation}
 H=\frac{p_\rho^2}{2} + \frac{p_z^2}{2} - \frac{1}{\sqrt{\rho^2{+}z^2}} + z\,F(t),
\end{equation}
where we have used atomic units.
The time evolution of the few-cycle pulse $F(t)$ with a Gaussian envelope is defined through the vector potential $A$
\begin{subequations}\label{eq:pulse}\begin{align}
F(t) &= -\frac{\partial }{\partial t}A(t)\\
A(t) &=-A_{0}\exp\left(-2\,{\rm ln}2
\left(t\big/nT\right)^{2}\right) \sin\left(\omega t+\phi\right),
\end{align}\end{subequations}
with $A_{0}=F_{0}/\omega$ the amplitude of the vector potential, $T=2\pi/\omega$ the laser period, $n$ the number of cycles, and $\phi$ the CEP.
Expression \eqref{eq:pulse} guarantees $\int{\rm d}t\,F(t)=0$.

\subsection{Initial conditions for trajectories}
The probability for tunneling at time $\ti$, where $\Fi=\left|F(\ti)\right|$, with a transverse momentum of $p_{\rho}(\ti)=\prhoi$ is given by \cite{shgo+09}
\begin{equation}
\label{eq:tunnelprobability}
W(\Fi,\prhoi)\propto\frac{1}{\Fi{}^2}{\rm exp}\left(-2\big(2E_{\rm ip}+\prhoi{}^2\big)^{3/2}\big/3\Fi\right)\frac{\prhoi}{\sqrt{1+\prhoi{}^2/2E_{\rm ip}}}
\end{equation}
with an additional Jacobian factor $\prhoi$.
$E_{\rm ip}$ is the atomic ionization potential.
In principle one could uniformly sample possible initial conditions for $\ti$ and $\prhoi$.
This, however, would be extremely inefficient due to the exponential weight in Eq.\,\eqref{eq:tunnelprobability}.
Most of the propagated trajectories would barely contribute due to their tiny weight.
Therefore, we define a grid of initial conditions with grid positions of $k_{i}=i\times\delta t$
where $|i|\le K/2$ in time and $l_{j}=j\times\delta p$ where $0\le j \le L$ in momentum.
The resulting $K{\times}L$ boxes in the ($\prhoi,\ti$) plane are now connected by some (arbitrary) path. 
With the definition
\begin{align}
X_{j} &\equiv \sum_{i=1}^{j}{\!}'\,W(|F(k_{i}{\times}\delta t)|,l_{i}{\times}\delta p),
\end{align}
where the prime indicates that the sum is calculated along the pre-defined path,
we get the probability
\begin{align}\label{eq:probability}
P_{j} &= X_{j}/X_{K{\times}L},
\end{align}
which is a monotonic function between 0 and 1 in terms of the index $j$.
Now we pick a uniform random number between 0 and 1, which corresponds by means of Eq.\,\eqref{eq:probability} to some $j$ and thus (again using the pre-defined path through the initial conditions) to $\ti=k_{j}{\times}\delta t$ and  $\prhoi=l_{j}{\times}\delta p$.
For the results presented below we have used $K={\rm int}(6nT/\delta t)$, $L=1000$, $\delta t=0.5$\,{a.u.} and $\delta p=0.001$\,a.u.

Electrons are propagated until $t=3nT$, where the laser field is sufficiently weak to be neglected. With the position and the momentum at this time one can easily calculate the asymptotic momenta $p_{z}''$ and $p_{\!\rho}''$.

\begin{figure}[b]
\begin{center}
\includegraphics[width=\textwidth]{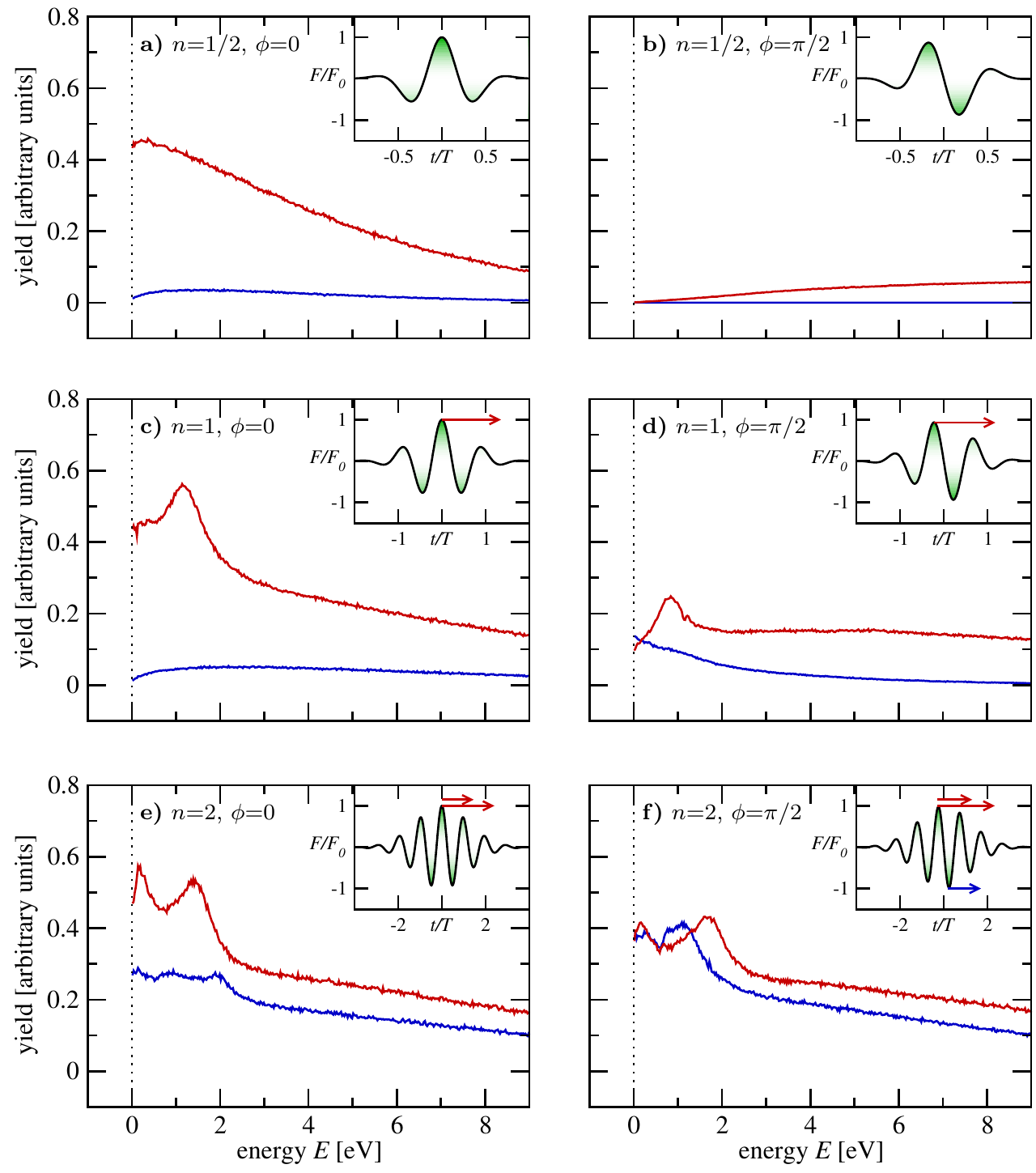}
\caption{Photo-electron spectra for energies $E=p''_{\!\rho}{}^{2}/2+p''_{z}{}^{2}/2$ of electrons detected upwards $\theta\,{<}\,5^{\circ}$ [red line] and downwards $\theta\,{>}\,175^{\circ}$ [blue line]
with $\tan(\theta)\equiv p''_{\!\rho}/p''_{z}$. 
Spectra are shown for pulses with $n=1/2$, 1, and 2 cycles [top to bottom row] and two CEPs $\phi=0$ [left column] and $\phi=\pi/2$ [right column], corresponding to sin-like and cos-like pulses, respectively.
The insets show the time-dependent electric field $F(t)$.
Each spectrum has been obtained by averaging over a Gaussian laser focus.}
\label{fig:spectra}
\end{center}
\end{figure}%

\section{Photo-electron spectra from few-cycle pulses}
In the following we present calculations for argon atoms ($E_{\rm ip}\,{=}\,0.5792$\,a.u.) and use laser pulses with a wavelength of $\lambda\,{=}\,2$\,$\mu$m ($\omega\,{=}\,0.0228$\,a.u.) and an intensity $I\,{=}\,10^{14}$W/cm$^2$ ($F_{0}\,{=}\,0.0534$\,a.u., $A_{0}\,{=}\,2.342$\,a.u.).
Thus the ponderomotive energy is $E_{\rm pond}\,{=}\,37.3$\,eV and the quiver amplitude $\zquiv\,{=}\,54.3$\,\AA.
Figure~\ref{fig:spectra} shows photo-electron spectra for pulses with an increasing number $n$ of cycles and two CEPs $\phi$ for each pulse length. 
The figure shows spectra obtained from averaging over intensities in a laser beam with a Gaussian focus \cite{aume+91}.

Not surprising, the most dramatic dependence on $\phi$ is seen for $n=1/2$ in Figs.\,\ref{fig:spectra}a and \ref{fig:spectra}b. 
The electron emission is preferentially in upward direction ($p''_{z}>0$).
Although this occurs for both CEPs the reason behind it is different in both cases.
The highest probability for tunneling is at the maxima of the electric field $F(t)$, cf.\ the green-shaded regions in the insets of Fig.\,\ref{fig:spectra}.
For $\phi=0$ there is one global maximum at $t_{\rm max}=0$. 
For a strong-field trajectory 
\begin{subequations}\label{eq:sfa-trajectory}\begin{align}
z(t,\ti) &= z'-A(t')\left[t-t'\right]+\int_{t'}^{t}{\rm d}\tau A(\tau)\\
p_{z}(t,\ti) &= A(t)-A(\ti),
\end{align}\end{subequations}
the drift momentum is given by $-A(t')$.
Thus for $n=1/2$ and $\phi=0$ one would expect an up-down symmetry since $A(t')$ is positive and negative for $t'\approx0$.
However, electrons, which are released towards negative $z$ (due to the negative electron charge a positive field corresponds to a force pointing in the negative direction), obtain an impact $\Delta p$ from the Coulomb potential which generates a positive offset for the drift momentum.
Therefore upward-emitted electrons are in the majority.
This holds for all pulse lengths (Figs.\,\ref{fig:spectra}a,\,c,\,e), whereby the difference gets weaker for longer pulses.
For a CEP of $\phi=\pi/2$ the energy distribution is shifted towards higher energies, see Fig.\,\ref{fig:spectra}b. 
Here the electric field has two maxima at $t_{\rm max}\approx\pm0.174T$.
Since the vector potential is negative for both times, namely $A(t_{\rm max})\approx-0.389A_{0}$, most of the electrons start with a positive drift momentum and the electrons are driven upwards.
The maximum of the distribution is expected around $E\approx0.302E_{\rm pond}=11.3$\,eV, i.\,e., outside the region shown in Figs.\,\ref{fig:spectra}b. 
Again the asymmetry\footnote{The CEP dependence of this asymmetry has recently been measured \cite{bezh+11}.} seen for the shortest pulse holds also for the longer pulses (Figs.\,\ref{fig:spectra}d,\,f), but becomes less and less distinct.

Most important, for the recollisions we are interested in, is that the spectra for $n=1/2$ are featureless at small energies.
This changes with $n=1$ where a low-energy peak shows up around $E\approx1$\,eV, see Figs.\,\ref{fig:spectra}c,\,d. 
Apparently now the pulse is sufficiently long, that the electron can be driven back at about $3/2$ cycles after its release. Figure \ref{fig:trajectories} shows a typical trajectory (solid green line), which is released at the maximum of the field ($t\approx0$) and returns to the nuclei, which we call a \emph{1st-order soft recollision}. 
Such trajectories are started at $t\approx0$ for $\phi=0$ (see inset of Fig.\,\ref{fig:spectra}c)
and at $t\approx-T/4$ for $\phi=\pi/2$ (see inset of Fig.\,\ref{fig:spectra}d).
The corresponding peak energies depend on $\phi$. 
This shows that the expression in Eq.\,(\ref{eq:epeaks}) should be used with care when applied to ultra-short pulses. In any case they give a good estimate for the peak locations. 

A further increase of the pulse duration shows for $n=2$ and $\phi=0$ (Fig.\,\ref{fig:spectra}e) a different spectrum.
For the first time one observes a double peak in the upward direction as well as a single peak for the downward-emitted electrons. This feature is much more pronounced for $\phi=\pi/2$ (Fig.\,\ref{fig:spectra}f).
Again, the arrows in the inset indicate the time span from tunneling to the recollision.
The longer red arrow marks a \emph{2nd-order soft recollision}, where the electron returns to the ion only after about $5/2$ cycles as shown by the dashed violet line in Fig.\,\ref{fig:trajectories}.  
Note that peaks occur for very small energies which makes it challenging for an experimental observation.

\section{Summary}
We have investigated theoretically low-energy photo-electron spectra generated by few-cycle long-wavelength laser pulses using a classical model.
A gradual increase of the pulse length supports the interpretation that the LES is due to soft recollisions with the ion.
Such recollisions can occur at later and later instances after the tunneling, rendering the LES  a series of peaks. However, these peaks can only emerge if the pulse has a sufficient number of cycles. 
Apart form the pulse length the photo-electron spectra depend also on the CEP. For a half-cycle pulse we predict that the LES vanishes.
An experimental realization of the proposed scenario should be feasible since CEP stabilization has been recently extended to the $\mu$m-wavelength regime \cite{scsh+11,bezh+11}.

\section*{References}
\def\harvardurl#1{\newline\texttt{#1}}\def\harvardurl#1{\relax}

\end{document}